\documentclass[9pt, sigconf]{acmart}
\usepackage{fancyhdr} 

\usepackage{svg}
\usepackage{array} 
\usepackage{pifont}
\usepackage{amsmath} 
\usepackage{threeparttable} 
\newcommand{\SetCaptionSpacing}[2]{%
    \setlength{\abovecaptionskip}{#1}%
    \setlength{\belowcaptionskip}{#2}%
}
\usepackage{multirow} 
\usepackage{multirow}
\usepackage{tikz}
\usepackage{graphicx}
\usepackage{cleveref}
\usepackage{csquotes}
\usepackage{etoolbox}
\usepackage{xcolor}
\usepackage{subfig}
\usepackage[table]{xcolor}
\usepackage[ruled,vlined,linesnumbered]{algorithm2e}
\usepackage{amsmath}
\usepackage{stfloats}

\definecolor{morandigreen}{HTML}{66CC00} 
\definecolor{morandigreen2}{HTML}{CCCC00}

\newcommand*\blackcircled[1]{%
    \tikz[baseline=(char.base)]{
        \node[shape=circle,
              draw,
              inner sep=0.2pt,
              minimum size=6pt,
              text=white,
              fill=black] (char) {#1};
    }
}
\definecolor{selfrowcolor}{RGB}{163,195,182}

\definecolor{darkgreen}{rgb}{0.0, 0.5, 0.0}

\def\BibTeX{{\rm B\kern-.05em{\sc i\kern-.025em b}\kern-.08em
    T\kern-.1667em\lower.7ex\hbox{E}\kern-.125emX}}

\copyrightyear{2026}
\acmYear{2026}
\setcopyright{cc}
\setcctype{by}
\acmConference[ICCAD '26]{IEEE/ACM International Conference on Computer-Aided Design}{November 08--12, 2026}{San Jose, CA, USA}
\acmBooktitle{IEEE/ACM International Conference on Computer-Aided Design (ICCAD '26), November 08--12, 2026, San Jose, CA, USA}
\acmDOI{10.1145/3831252.3834064}
\acmISBN{979-8-4007-2873-0/2026/11}

\begin{document}
\title{
    E-ALS: Approximation-Potential-Aware E-Graph Rewriting for Approximate Logic Synthesis
}


\author{
  Bin Sun$^{1,2}$,
  Jianan Mu$^{1}$,
  Jiaxi Zhang$^{4}$,
  Rengang Zhang$^{1,2}$, 
  Zhiteng Chao$^{1}$,
  Jing Ye$^{1,2,3}$,
  Huawei Li$^{1,2,3}$
}

\thanks{
    Corresponding authors:\{mujianan,chaozhiteng\}@ict.ac.cn,zhangjiaxi@pku.edu.cn. \\
    This work was supported by the National Natural Science Foundation of China (Grant Nos. 92473203, 92373206, and U25A20486) and the Strategic Priority Research Program of Chinese Academy of Sciences (Grant No. XDB0660103).
}

\affiliation{%
  \institution{%
    $^1$State Key Lab of Processors, Institute of Computing Technology, CAS, Beijing, China \\
    $^2$University of Chinese Academy of Sciences, Beijing, China \  $^3$CASTEST Co., Ltd. \\
    $^4$Department of Computer Science, Peking University, Beijing, China 
  }
  \country{} 
}


\renewcommand{\authors}{
  Bin Sun, Jianan Mu, Jiaxi Zhang, Rengang Zhang,
  Zhiteng Chao, Jing Ye, Huawei Li
}

\renewcommand{\shortauthors}{Sun et al.}


\begin{abstract}
Approximate logic synthesis (ALS) improves circuit power, performance, and area by trading exact correctness for bounded functional error. 
However, existing structural ALS methods largely overlook \emph{structural bias}: even functionally equivalent netlists can expose markedly different approximation opportunities and lead to substantially different outcomes under the same downstream ALS flow. Our experiments show that this effect can induce final area gaps of up to 42.77\%.
To unlock this opportunity, we propose \textbf{E-ALS}, an e-graph-based framework for approximation-aware structural search. 
E-ALS combines Function-Reduced Saturation, an ALS-coupled surrogate, search-based extraction, and budget-guided refinement to identify approximation-friendly equivalent structures. 
Experiments on well-established arithmetic and logic benchmarks show that \textbf{E-ALS} achieves additional area reductions of 3.2 and 7.0 percentage points under maximum Hamming-Distance and Error-Distance constraints, respectively.
Code is available in \url{https://github.com/ZenuSunB/Ecompile.git}
\end{abstract}




\maketitle
\section{Introduction}
\label{intro}

The growing prevalence of error-resilient applications, such as neural inference and large language model (LLM) workloads, has renewed interest in approximate logic synthesis (ALS). By trading bounded functional error for improvements in power, performance, and area (PPA), ALS offers an attractive opportunity for hardware optimization in error-tolerant computing systems~\cite{scarabottolo2020approximate,shin2010approximate}.

Existing ALS techniques can be broadly divided into functional and structural methods~\cite{scarabottolo2020approximate}. Functional methods directly modify the Boolean function but often face scalability challenges. Structural methods instead operate through local approximate changes (LACs) on the logic netlist and therefore scale more effectively to large designs. Along this direction, increasingly expressive approximation operators, ranging from const-LAC and SASIMI-style substitutions~\cite{venkataramani2013substitute} to resubstitution-based transformations~\cite{meng2024efficient,meng2020alsrac} and approximate subcircuit replacement~\cite{rezaalipour2025approximate}, have substantially improved ALS capability. However, these efforts mainly enlarge the approximation operator itself, rather than systematically selecting the exact logic on which structural ALS is applied.

\begin{figure}[t]
  \centering
  \SetCaptionSpacing{0.10cm}{0.10cm}
  \subfloat[]{ \includegraphics[width=0.47\linewidth]{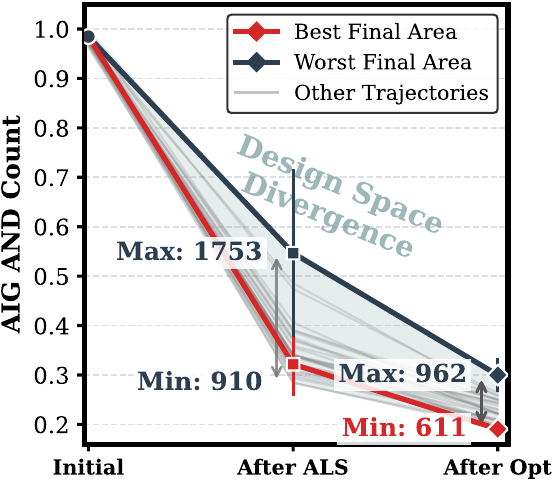}\label{fig:problem-a}}
  \subfloat[]{ \includegraphics[width=0.52\linewidth]{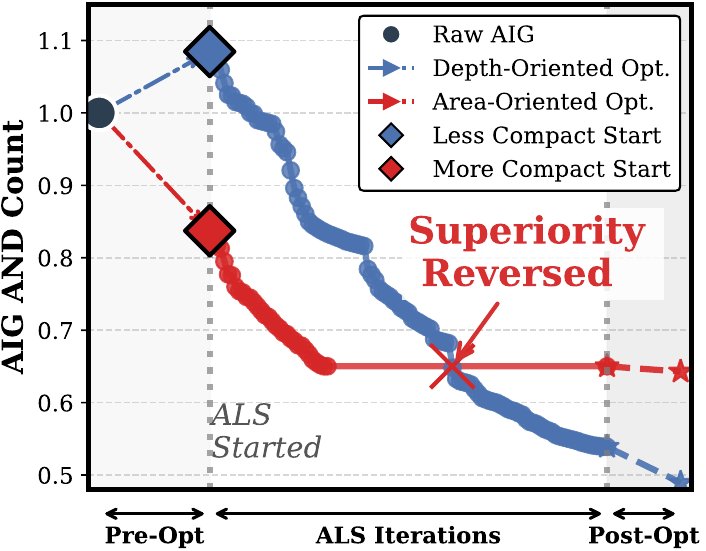}\label{fig:problem-b}}
  \caption{(a) Functionally equivalent starting structures yield divergent ALS trajectories and final results. (b) A structure superior under exact optimization is not necessarily optimal for downstream approximation.}
  \vspace{-8mm}
\end{figure}

This distinction is important because, in structural ALS, approximation operators are inherently structure-dependent: even functionally equivalent netlists may expose markedly different local approximation opportunities and error-propagation patterns, leading to substantially different approximation outcomes. Since structural ALS is typically guided by heuristic optimization rather than global search, this dependence is further amplified. We refer to this phenomenon as \emph{structural bias}: under heuristic structural ALS, different but functionally equivalent logic representations can expose systematically different approximation opportunities to the synthesis flow. As illustrated in Fig.~\ref{fig:problem-a} and Fig.~\ref{fig:problem-b}, structural bias manifests in two ways: functionally equivalent starting structures can lead to different ALS trajectories and final results, and a representation that is superior under exact optimization is not necessarily more favorable for approximation. In other words, a netlist that appears more compact or better optimized before approximation does not necessarily yield a better approximate implementation.

These observations suggest that, rather than performing ALS on a single exact netlist, one should optimize over a space of functionally equivalent structures to uncover approximation opportunities hidden by any single topology. E-graphs provide a compelling substrate for this purpose~\cite{nandi2021rewrite,willsey2021egg}, since they can compactly preserve many equivalent structures without explicit enumeration. However, for structural ALS, preserving a large equivalence space alone is not enough: the space must also be usable for downstream search and selection.

This gives rise to two challenges. \textbf{First, how can one construct a structural search space that is not only rich in equivalent implementations, but also consolidated and exploitable for downstream search?} Since the equivalence space is astronomically large, naive rewriting may enlarge the space without aiding approximation-oriented extraction. \textbf{Second, how can one efficiently evaluate candidates during structure search according to their downstream approximation quality?} A structure that appears superior under exact synthesis metrics does not necessarily lead to a better approximate result, so conventional exact-area criteria cannot reliably guide selection. Meanwhile, evaluating many equivalent structures with the full exact legality-checking flow would be prohibitively expensive. This calls for an efficient ALS-coupled surrogate that can rank candidate structures without repeatedly invoking expensive exact verification.

To this end, we propose \textbf{E-ALS}, an e-graph-based framework for approximate logic synthesis. E-ALS is built on two core components. The first is \emph{equivalence-enhanced structural search space construction}, which combines equivalence rewriting with \emph{Function-Reduced Saturation} to construct a richer yet more consolidated equivalence space for downstream approximation. The second is an \emph{ALS-coupled surrogate}, which provides an efficient approximation-potential-aware objective for ranking candidate structures during search. Built upon these two components, E-ALS performs search-based extraction of promising topologies and further applies budget-guided structural refinement under progressively relaxed error bounds.




\textbf{Our contributions are summarized as follows:}
\begin{itemize}
\item We systematically demonstrate that structural ALS performance depends strongly on logic representation, and identify this effect as \emph{structural bias} in heuristic ALS flows.

\item We propose \textbf{E-ALS}, an e-graph-based framework that explores diverse equivalent logic structures and evaluates their approximation potential through ALS-coupled evaluation, search-based extraction, and budget-guided refinement.

\item Experiments on arithmetic and logic benchmarks show that \textbf{E-ALS} achieves average normalized final areas of 54.94\% and 74.69\% under maximum Hamming Distance and Error Distance constraints, reducing area by an additional 3.2 and 7.0 percentage points over representative baselines.
\end{itemize}

\section{Background and Related Work}
\label{relatework}

\subsection{Structural Approximate Logic Synthesis}
Structural approximate logic synthesis (ALS) improves scalability by introducing bounded functional errors~\cite{scarabottolo2020approximate}. We focus on the maximum-error constraint, which guarantees explicit worst-case bounds. Prior works enforce this through approximate don't-cares (SALSA~\cite{venkataramani2012salsa}), minimal unsatisfiable subsets (MUSCAT~\cite{witschen2022muscat}), partial Boolean differences (MECALS~\cite{meng2023mecals}), simulation-guided SAT solving~\cite{meng2026simulation}, and SMT-based subcircuit replacement (SubXPAT~\cite{rezaalipour2025approximate}). Open resources like EvoApproxLib~\cite{mrazek2019evoapproxlib} further aid such research.

To improve final approximate \emph{quality}, recent efforts pursue two paradigms. The first enhances local approximate change (LAC) to expose richer opportunities, utilizing substitute-and-simplify transformations~\cite{venkataramani2013substitute}, node simplification~\cite{wu2019alfans}, resubstitution~\cite{meng2020alsrac,meng2024efficient}, and subcircuit replacement~\cite{rezaalipour2025approximate}. The second optimizes the heuristic flows surrounding these operators, leveraging sensitivity-driven screening~\cite{meng2022seals}, simultaneous LAC application~\cite{wang2023accals,wang2024accals}, differentiable search~\cite{wang2023dasals}, evolutionary optimization~\cite{balaskas2022variability}, and tree search~\cite{ye2025rank,hu2025timing}. While these advances substantially refine \emph{how} approximation is performed on a given structure, they overlook a critical question: among functionally equivalent exact structures, which one provides the optimal substrate for downstream structural ALS?

\subsection{Equivalence-Based Structural Exploration}
The impact of logic structure is well-recognized in exact synthesis. While earlier studies identified structural bias in technology mapping~\cite{chatterjee2006reducing}, their structural preservation targeted exact metrics (e.g., area, delay) rather than approximation quality. Recently, e-graph-based methods have emerged as a powerful tool to compactly preserve and explore equivalent structures for logic synthesis (E-Syn~\cite{chen2024syn}, E-morphic~\cite{chen2025morphic}), FPGA remapping (EqMap~\cite{hofmann2025eqmap}), and technology mapping~\cite{chen2025revisit,yin2025boost}. These works prove the efficacy of equality saturation for exact optimization. Our work uniquely bridges this capability into the ALS domain. Unlike exact synthesis, our objective is not immediate structural compactness, but rather identifying equivalent exact topologies that unlock high downstream approximation potential under a maximum-error constraint.

\section{Empirical Observations and Implications}
\label{motivation}

\textbf{OBSERVATION 1: The initial logic structure strongly affects ALS outcomes.} 
To evaluate this effect, we generated structurally diverse, functionally equivalent netlists for each benchmark using e-graph rewriting and randomized extraction, and applied the same ALS flow~\cite{meng2026simulation} to each variant. Each experiment was repeated three times, and a uniform \texttt{"2$\times$compress2rs"} post-processing step was applied to reduce bias from inconsistent exact cleanup.

\begin{figure}[tb!]
  \centering
  \SetCaptionSpacing{0.10cm}{0.10cm}
  \includegraphics[width=1.0\linewidth]{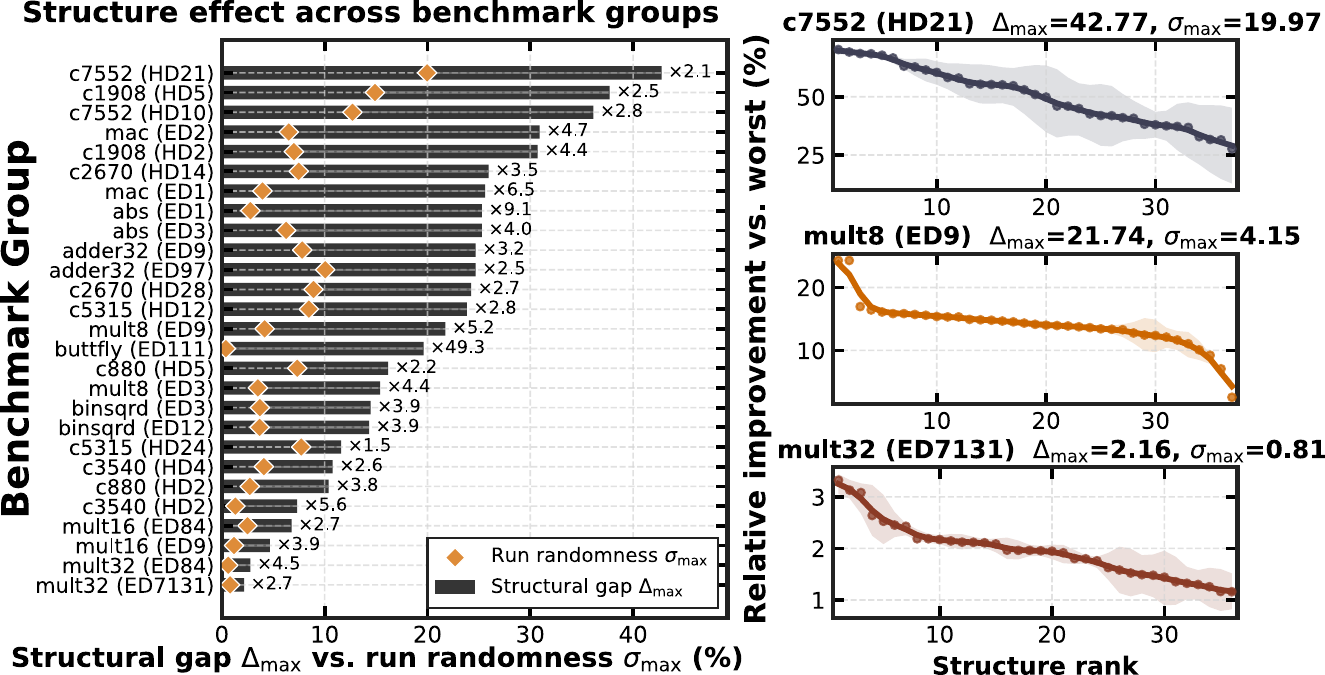}
  \caption{\textbf{Structural effect vs. algorithmic randomness}. Comparison between structural effect ($\Delta_{\max}$) and algorithmic randomness ($\sigma_{\max}$), indicating that structural choice is a dominant factor in ALS quality.}
  \label{fig:motivation1}
  \vspace{-5mm}
\end{figure}

Fig.~\ref{fig:motivation1} shows that different equivalent structures can lead to substantially different post-ALS results. In the most extreme case, \texttt{c7552(HD21)}, the gap between the best and worst mean outcomes reaches 42.77\%; across many benchmark groups, the gap is still around 10\%--30\%. More importantly, this structural effect consistently exceeds run-to-run randomness: $\Delta_{\max}$ is larger than $\sigma_{\max}$ in all benchmark groups, e.g., 42.77\% vs.\ 19.97\% for \texttt{c7552(HD21)} and 21.74\% vs.\ 4.15\% for \texttt{mult8(ED9)}. Therefore, the input structure is not merely a neutral starting point, but a key factor affecting downstream ALS quality.

\textbf{IMPLICATION.} Effective structural ALS should preserve and explore rich structural diversity among functionally equivalent implementations, rather than relying on a single exact topology.

\begin{figure}[tb!]
  \centering
  \SetCaptionSpacing{0.10cm}{0.10cm}
  \includegraphics[width=1.0\linewidth]{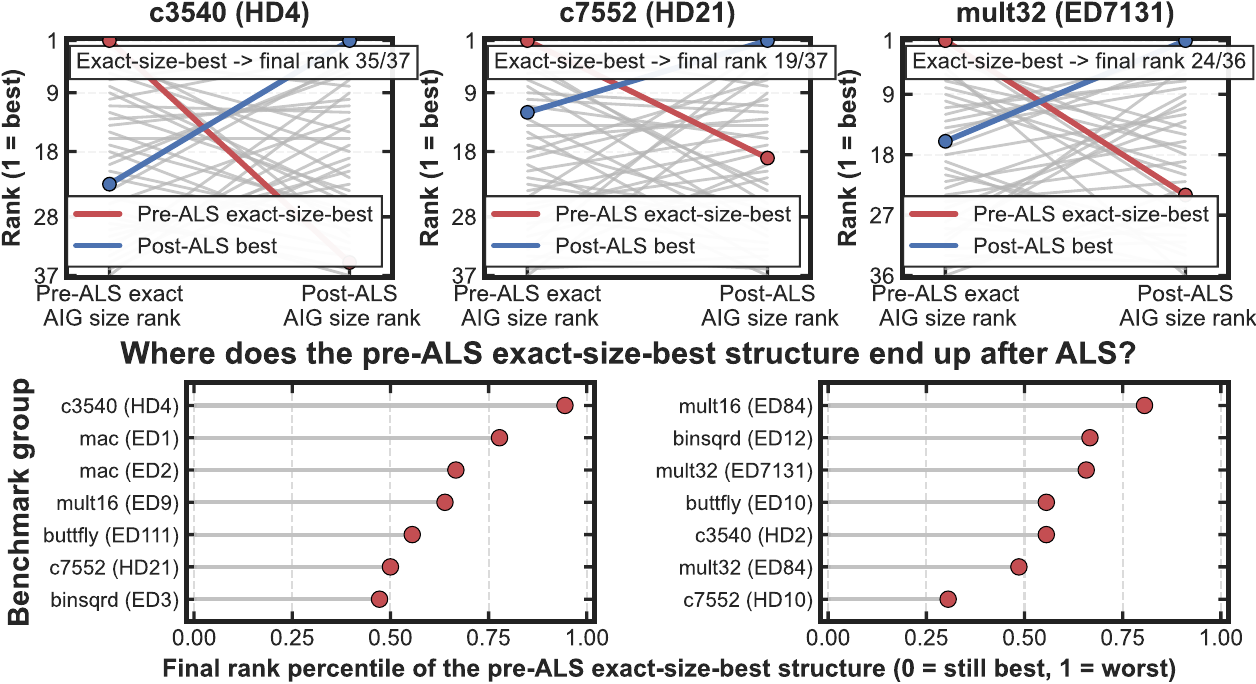}
  \caption{\textbf{Suboptimality of the exact-size-best structure.} The structure with the smallest exact AIG size before ALS is often not the best after ALS.
  }
  \label{fig:motivation2}
  \vspace{-7mm}
\end{figure}

\textbf{OBSERVATION 2: Exact pre-ALS compactness does not reliably identify the best starting structure.}
A natural strategy is to choose the most compact exact structure before ALS, where compactness is measured by exact AIG size. However, Fig.~\ref{fig:motivation2} shows that this strategy is often misleading: the exact-size-best candidate frequently ranks only in the middle or even near the bottom after ALS. For example, it ranks 35th out of 37 for \texttt{c3540(HD4)}, 19th out of 37 for \texttt{c7552(HD21)}, and 24th out of 36 for \texttt{mult32(ED7131)}.
This mismatch indicates that ALS-friendliness is not determined by exact compactness alone. What matters is whether a structure exposes favorable approximation opportunities and error-propagation patterns for the downstream ALS flow. Hence, exact pre-ALS compactness is insufficient for structure ranking.

\textbf{IMPLICATION.} Structural search requires an approximation-aware evaluation mechanism that estimates the downstream approximation potential of candidate structures, rather than relying solely on exact pre-ALS compactness.

These observations suggest two requirements for E-ALS: preserving a rich space of functionally equivalent structures, and evaluating them according to their downstream approximation potential. These motivate the equivalence-enhanced structural search in Section~4.2 and the ALS-coupled surrogate evaluation in Section~4.3.

\section{Methodology and Design of E-ALS}
\label{Methods}

\begin{figure*}[t]
  \centering
  \SetCaptionSpacing{0.10cm}{0.10cm}
  \includegraphics[width=0.9\linewidth]{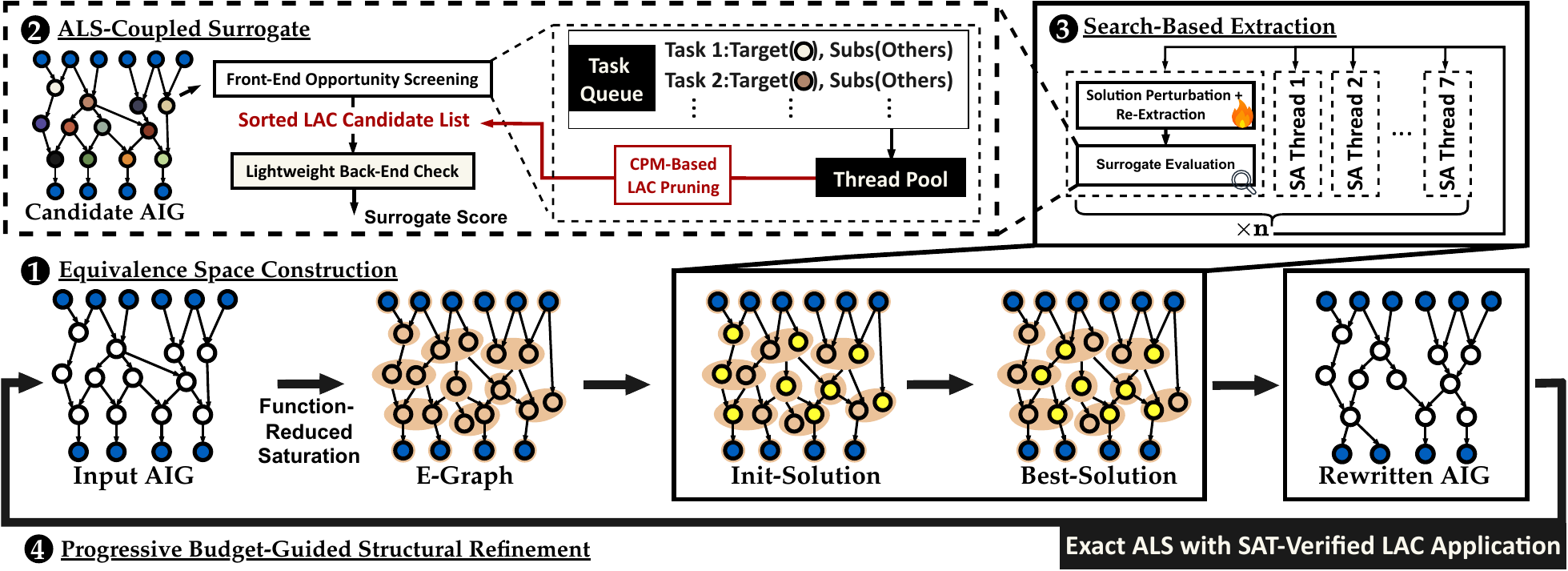}
  \caption{Framework of E-ALS.}
  \label{fig:framework}
  \vspace{-6mm}
\end{figure*}

\subsection{Problem Formulation and Overview}
\label{sec:problem}

Guided by the observations in Section~3, we formulate structural ALS as a structure-selection problem over the equivalence space of the input netlist. Given an exact netlist $N_0$, an error constraint $\epsilon$, and an ALS algorithm $\mathcal{A}$, we seek an equivalent structure that minimizes the post-ALS area. Using post-ALS AIG size as a tractable proxy, the objective is
\begin{equation}
N^{*} = \mathop{\arg\min}\limits_{N \in \mathcal{S}(N_0)} \; \mathrm{Size}_{\mathrm{AIG}}\!\left(\mathcal{A}(N,\epsilon)\right),
\label{eq:problem}
\end{equation}
where $\mathrm{Size}_{\mathrm{AIG}}(\cdot)$ denotes the AIG node count of the resulting netlist.

This formulation requires both a rich yet manageable representation of equivalent structures and an effective measure of their \textbf{\emph{approximation potential}}, since exact pre-ALS compactness does not reliably predict downstream quality. E-ALS combines \emph{equivalence-enhanced structural search space construction} (Section~4.2), based on equivalence rewriting and \emph{Function-Reduced Saturation}, with \emph{ALS-coupled surrogate evaluation} (Section~4.3) to estimate the achievable post-ALS AIG size. It further employs \emph{search-based extraction} (Section~4.4) to obtain promising structures and \emph{progressive budget-guided refinement} (Section~4.5) to revisit structure selection only when an enlarged error budget may change candidate rankings.

As illustrated by Fig.~\ref{fig:framework}~\blackcircled{1}$\sim$\blackcircled{4}, E-ALS first constructs an enriched e-graph that compactly preserves diverse equivalent topologies, evaluates candidate structures with the ALS-coupled surrogate objective, extracts promising structures for downstream ALS, and progressively refines the selection as the error budget increases.

\subsection{Equivalence-Enhanced Structural Search Space Construction}

Step \blackcircled{1} addresses the first core requirement of E-ALS: constructing a structural search space that is rich enough to expose alternative implementations, yet consolidated enough to remain exploitable during extraction. To this end, we build an e-graph over the exact input network using an extended AIG representation. By explicitly incorporating \texttt{OR} nodes alongside \texttt{AND} and \texttt{INV}, we shorten Boolean rewrite distances among equivalent structures and enlarge the reachable topology space, thereby improving the structural diversity available for downstream ALS.

However, naive saturation alone is insufficient to meet this requirement. Although rewriting can expand the topology pool, it often suffers from \emph{semantic fragmentation}: functionally equivalent but syntactically distinct structures remain isolated in disjoint e-classes, so the search space becomes larger without becoming correspondingly more exploitable for downstream extraction.

To address this issue, we propose \textbf{Function-Reduced Saturation (FRS)} (Algorithm~\ref{alg:function_reduced_saturation}), which alternates node-limited saturation with periodic functional consolidation. Here, node-limited saturation controls e-graph growth while preserving structural expansion, and functional consolidation merges semantically equivalent regions to reduce fragmentation. After each saturation, \textsc{FunctionReduceCheck} identifies functionally equivalent regions, and \textsc{ApplyFunctionReduceUnion} merges the corresponding e-classes. In this way, rewriting expands the structural space, while function reduction consolidates semantic equivalence, making the resulting e-graph not merely larger, but denser and more exploitable for downstream search.

\begingroup
\SetAlCapNameFnt{\footnotesize\bfseries}
\SetAlCapFnt{\footnotesize}
\begin{algorithm}[t]
\caption{FRS for Structural Search Space Construction}
\label{alg:function_reduced_saturation}
\fontsize{7.5}{8.0}\selectfont
\KwIn{$\mathcal{G}_0$: initial e-graph; $\mathcal{R}$: rewrite rule set;
      $\rho$: max node-expansion-ratio; $R$: number of rounds}
\KwOut{$\mathcal{G}_{\mathrm{sat}}$: saturated e-graph}
$\mathcal{G} \leftarrow \mathcal{G}_0$\,,\quad
$M \leftarrow \rho \, |\mathcal{G}_0|$\;
\For{$r = 1$ \KwTo $R$}{
    $M_r \leftarrow
    \left\lfloor \dfrac{r}{R} M \right\rfloor$\;
    $\mathcal{G} \leftarrow
    \textsc{NodeLimitedSaturation}(\mathcal{G},\mathcal{R},M_r)$\;
    $\Pi_r \leftarrow
    \textsc{FunctionReduceCheck}(\mathcal{G})$\;
    $\mathcal{G} \leftarrow
    \textsc{ApplyFunctionReduceUnion}(\mathcal{G},\Pi_r)$\;
    \lIf{$|\mathcal{G}| \ge M$}{\textbf{break}}
}
\Return{$\mathcal{G}_{\mathrm{sat}} \leftarrow \mathcal{G}$}\;
\end{algorithm}
\endgroup

\begin{figure}[tb!]
  \centering
  \SetCaptionSpacing{0.10cm}{0.10cm}
  \includegraphics[width=0.62\linewidth]{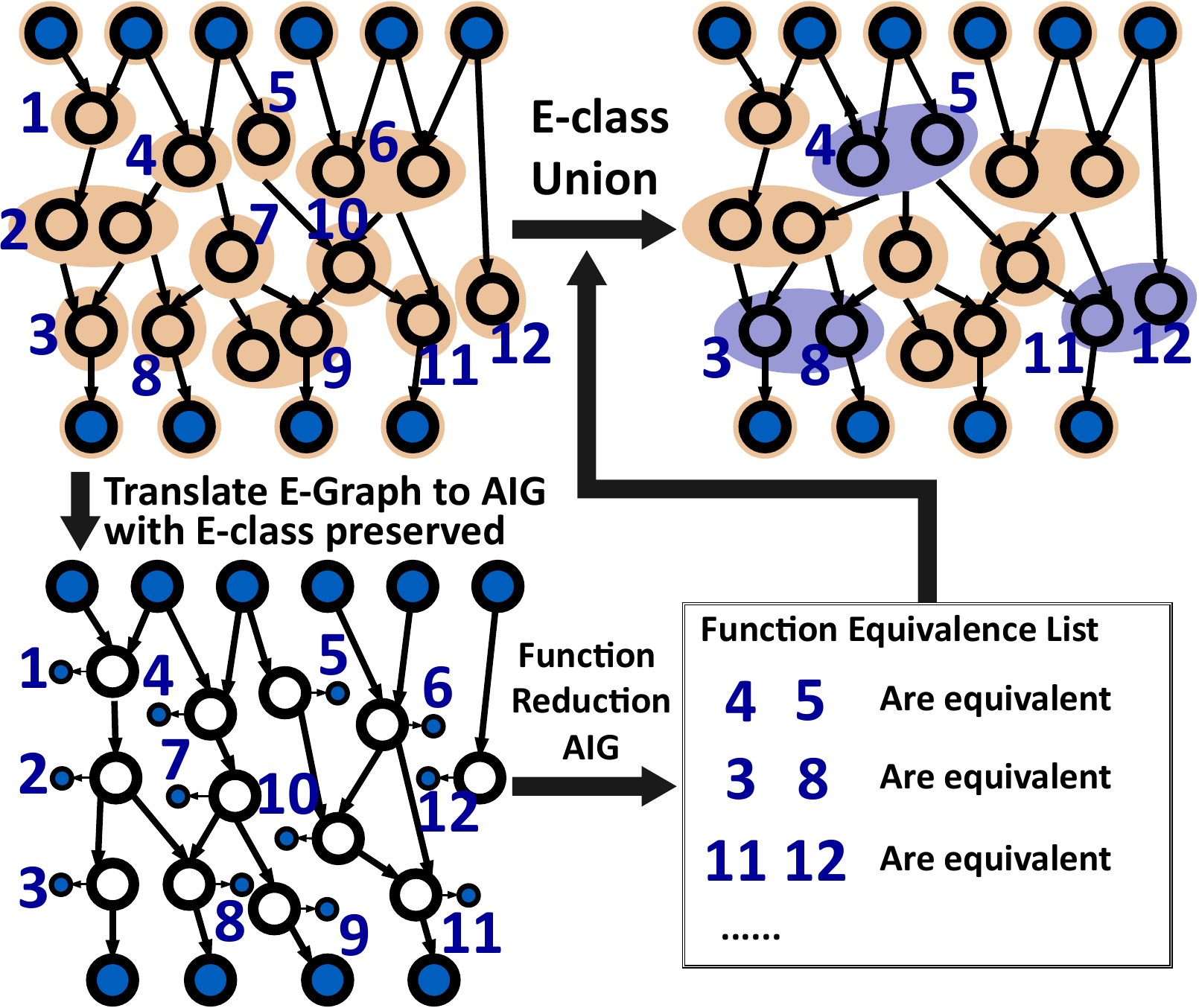}
  \caption{Function reduction in the e-graph via AIG-based equivalence checking.}
  \label{fig:method1}
  \vspace{-6mm}
\end{figure}

As illustrated in Fig.~\ref{fig:method1}, \textsc{FunctionReduceCheck} is implemented by translating the current e-graph into an AIG while preserving the mapping from each exported AIG node back to its source e-class. \texttt{ABC}~\cite{brayton2010abc} \texttt{"fraig"} is then applied to detect functionally equivalent AIG nodes, producing equivalence pairs such as $(4,5)$, $(3,8)$, and $(11,12)$ in Fig.~\ref{fig:method1}. Through the preserved ID mapping, these equivalences are projected back to the original e-graph, where \textsc{ApplyFunctionReduceUnion} unions the corresponding e-classes. In this way, exact functional equivalence discovered in the AIG domain is reinjected into the e-graph as e-class unions, enabling semantic consolidation during saturation.

\subsection{ALS-Coupled Surrogate Evaluation of Approximation Potential}

Step \blackcircled{2}addresses the second core requirement of E-ALS: candidate structures cannot be ranked by exact pre-ALS compactness alone, while evaluating every structure with the full exact ALS flow is too expensive. We therefore develop an \emph{ALS-coupled surrogate} that preserves the structure-sensitive front end of ALS and replaces the expensive exact back-end legality checking with a lightweight simulation-based alternative. In E-ALS, under a given error bound, the approximation potential of a candidate structure is approximated by the post-ALS AIG size estimated by this surrogate.

The surrogate consists of two parts: front-end opportunity screening and back-end lightweight structure evaluation.

\subsubsection{Front-End Opportunity Screening}

We retain the front end of conventional maximum-error ALS because it provides a structure-sensitive signal on how many promising local approximation opportunities a candidate netlist can expose. Given a candidate exact netlist, we enumerate SASIMI LACs~\cite{venkataramani2013substitute} and use change propagation matrix (CPM)-based simulation pruning to discard clearly infeasible candidates~\cite{su2022vecbee}.

Following the CPM-based screening logic in~\cite{su2022vecbee}, we evaluate each candidate LAC
\(
\ell=(t,s,\sigma)
\)
over a sampled pattern set
\(
\mathcal{X}=\{x^{(1)},\dots,x^{(M)}\}
\).
Here, \(t\) is the target node, \(s\) is the substitute node, and \(\sigma\) denotes optional inversion. Let
\(
\delta_{\ell}^{(i)} = v_t^{(i)} \oplus v_{s,\sigma}^{(i)}
\)
denote whether the substitute differs from the original target under pattern \(x^{(i)}\), where \(v_{s,\sigma}^{(i)}\) is the simulated value of \(s\) after applying \(\sigma\). We use the Boolean change-propagation matrix \(P(i,t,o)\) to indicate whether flipping target \(t\) changes output \(o\) under \(x^{(i)}\). The induced output-error mask is therefore
\(
\delta_{\ell}^{(i)} \wedge P(i,t,:)
\),
which yields the simulation-based lower bound
\begin{equation}
\widehat{E}_{\ell}^{\,\mathrm{LB}}
=
\max_{x^{(i)}\in\mathcal{X}}
D\!\left(
y^{(i)},
\, y^{(i)} \oplus \bigl(\delta_{\ell}^{(i)} \wedge P(i,t,:)\bigr)
\right),
\label{eq:lb_error}
\end{equation}
where \(D(\cdot,\cdot)\) is the chosen error metric (e.g., HD/ED). Any candidate with
\(
\widehat{E}_{\ell}^{\,\mathrm{LB}} > \epsilon
\)
is pruned, where \(\epsilon\) is the maximum-error bound.

The key computational reuse in this screening is target-centric. For all LACs sharing the same target \(t\), the expensive propagation context is identical, while different substitutes affect only the local discrepancy \(\delta_{\ell}^{(i)}\). Equivalently, these candidates share the same CPM slice
\(
\mathbf{P}_t = P(:,t,:)\in\{0,1\}^{M\times O}
\).
We therefore group all substitutes of the same target into one evaluation task, construct \(\mathbf{P}_t\) once, and reuse it across the entire substitute set. In implementation, targets are further packed into fixed-size blocks of 16 and scheduled through a load-balanced task queue, so targets with unusually large substitute sets or heavier local cones do not become bottlenecks. This target-centric organization preserves the structural sensitivity of the ALS front end while making raw-LAC screening efficient enough to serve as the first part of our ALS-coupled surrogate.

\subsubsection{Back-End Lightweight Structure Evaluation}

The conventional back end of maximum-error ALS greedily applies the promising LACs that survive front-end pruning and performs exact SAT-based legality checking after each application. Repeating this exact process for many candidate structures is prohibitively expensive. We therefore retain the same greedy apply-and-check skeleton, but replace exact legality checking with lightweight simulation.

\begin{figure}[tb!]
  \centering
  \SetCaptionSpacing{0.10cm}{0.10cm}
  \includegraphics[width=0.75\linewidth]{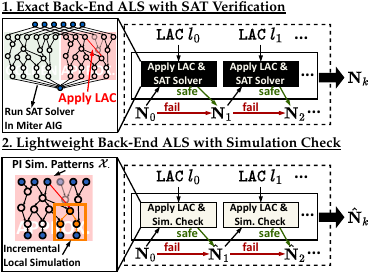}
  \caption{Exact versus lightweight back-end ALS for candidate-structure evaluation.}
  \label{fig:method2}
  \vspace{-6mm}
\end{figure}

This replacement is efficient because a SASIMI LAC~\cite{venkataramani2013substitute} modifies only one target driver and its downstream fanout region. As illustrated in Fig.~\ref{fig:method2}, instead of rebuilding and verifying the whole circuit after each LAC, we locally apply the LAC, incrementally simulate only the affected region, and accept or reject the modification using a lightweight error check.

Concretely, let \(\hat{N}_j\) denote the current lightweight approximate circuit at step \(j\), and let \(\ell_j\) be the next promising LAC selected by the front end. Applying \(\ell_j\) produces a temporary locally modified circuit \(\hat{N}_j^{(\ell_j)}\). We then perform an incremental simulation-based error check on \(\hat{N}_j^{(\ell_j)}\) over \(\mathcal{X}\):
\begin{equation}
\widehat{E}_{\mathrm{sim}}\!\left(\hat{N}_j^{(\ell_j)}\right)
=
\mathrm{SimCheck}\!\left(\hat{N}_j^{(\ell_j)}, \mathcal{X}\right),
\label{eq:backend_simcheck}
\end{equation}
where \(\mathrm{SimCheck}(\cdot)\) returns a lightweight estimate of the induced error. If the modified circuit passes this check, the LAC is accepted and
\begin{equation}
\hat{N}_{j+1} = \hat{N}_j^{(\ell_j)};
\label{eq:backend_accept}
\end{equation}
otherwise, the local patch is rolled back and
\begin{equation}
\hat{N}_{j+1} = \hat{N}_j.
\label{eq:backend_reject}
\end{equation}

Repeating this process yields a lightweight approximate trajectory over \emph{accepted} intermediate circuits,
\begin{equation}
\hat{N}_0 \rightarrow \hat{N}_1 \rightarrow \cdots \rightarrow \hat{N}_K,
\label{eq:lightweight_traj}
\end{equation}
where each \(\hat{N}_j\) denotes the current lightweight approximate circuit after the \(j\)-th accepted LAC application. Rejected LACs are rolled back immediately and therefore do not advance the trajectory. We define
\begin{equation}
K=\min \left\{ j \ge 1 \;\middle|\; \widehat{E}_{\mathrm{sim}}(\hat{N}_j)=\epsilon \right\},
\label{eq:surrogate_stop}
\end{equation}
that is, \(K\) is the first accepted step at which the current lightweight approximate circuit reaches the error budget \(\epsilon\) under simulation. This stopping rule reduces the optimism of the surrogate by terminating the lightweight trajectory immediately when the error budget is first reached, rather than continuing to scan additional LACs along the budget boundary.
We then define the surrogate score of candidate structure \(N\) as
\begin{equation}
\widehat{S}_{\mathrm{AIG}}(N) = \mathrm{Size}_{\mathrm{AIG}}(\hat{N}_K).
\label{eq:surrogate_score}
\end{equation}
A smaller \(\widehat{S}_{\mathrm{AIG}}(N)\) indicates a more favorable starting structure for downstream ALS.

\subsection{Search-Based Extraction of High-Potential Structures}

\begin{figure}[t]
  \centering
  \SetCaptionSpacing{0.10cm}{0.10cm}
  \includegraphics[width=1\linewidth]{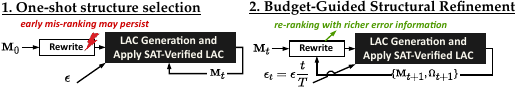}
  \caption{One-shot structure selection versus budget-guided structural refinement. Here, $M_t$ denotes the current circuit entering budget stage $t$, and $\Omega_t$ denotes the accumulated counterexample patterns available at that stage.}
  \label{fig:method3}
  \vspace{-6mm}
\end{figure}

With the equivalence space from Section~4.2 and the surrogate objective from Section~4.3 in place, the search-based extraction procedure (\blackcircled{3}in Fig.~\ref{fig:framework}) identifies a concrete structure for downstream ALS. Since the objective is defined by surrogate-estimated post-ALS AIG size rather than by a simple exact metric, we adopt simulated annealing~\cite{kirkpatrick1983optimization} as a practical search mechanism over extractable topologies.

The search is initialized from the original AIG \(N_{\mathrm{orig}}\), avoiding early commitment to an exact-oriented structural prior. At each iteration, a batch of neighboring candidates is generated around the current solution. Each candidate is first obtained through \emph{perturbation and re-extraction}~\cite{chen2025morphic}, which partially perturbs the current structure and incrementally re-extracts the affected region, and is then further processed by a \emph{fast refinement} step that applies one randomly selected short exact-oriented ABC optimization script (e.g., \texttt{balance}, \texttt{rewrite}, \texttt{refactor}, \texttt{resub}, and related short flows), followed by a fixed post-processing step, to generate additional nearby variants. All candidates are scored by the surrogate objective, and the best one in the batch is selected as the proposal for acceptance. Simulated annealing then decides whether this proposal replaces the current solution, while the best solution found so far is maintained throughout the search. In this way, the extraction procedure turns the enriched equivalence space into a searchable domain and guides the search toward structures that are more favorable for downstream ALS.



\subsection{Budget-Guided Structural Refinement}

As discussed in Section~4.3, E-ALS ranks structures using a lightweight ALS-coupled surrogate rather than repeated exact legality checking. While this makes large-scale structural search practical, it also makes one-shot structure selection brittle: an early surrogate mis-ranking may place the downstream ALS flow on a suboptimal trajectory, and the resulting quality gap can persist throughout later optimization, as illustrated on the left of Fig.~\ref{fig:method3}.

Our key observation is that structural re-ranking becomes more reliable when more accumulated counterexample patterns $\Omega_t$ are available and the stage-wise error budget is closer to the final target bound. We therefore perform structural refinement under a progressive budget schedule
\[
\epsilon_t = \left\lceil \frac{t}{T}\epsilon \right\rceil,\qquad t=1,2,\ldots,T .
\]
At stage $t$, E-ALS uses the current circuit $M_t$, the accumulated counterexample set $\Omega_t$, and the current bound $\epsilon_t$ to re-rank candidate structures and re-extract a favorable one, and then performs ALS under the same bound before proceeding to the next stage. In this way, structure selection is revisited only when the feasible approximation region has expanded enough to potentially change the relative ranking of candidate structures.

\section{Experiments}\label{experiments}
\subsection{Experimental Setup}

We evaluate E-ALS on the benchmarks in Table~\ref{tab:experimental_benchmarks}, including ISCAS85~\cite{hansen2002unveiling}, the EPFL arithmetic suite~\cite{amaru2015epfl}, and circuits used in prior MECALS evaluation~\cite{meng2023mecals}. For each benchmark, we consider three exact structural representations: a \emph{naive AIG} (\texttt{"strash"}), a \emph{compact AIG} (\texttt{"2$\times$compress2rs"}), and a \emph{flattened AIG} (\texttt{"2$\times$if-K6-g-C8"}). Table~\ref{tab:experimental_benchmarks} summarizes their statistics, together with mapped gate-level area and delay in the ASAP 7\,nm library~\cite{vashishtha2017asap7}.

We consider both Hamming-distance (HD) and error-distance (ED) constraints. For HD, the MaxHD bounds are set to $\lfloor 0.1O \rfloor$ and $\lfloor 0.2O \rfloor$ for each benchmark, where $O$ is the number of primary outputs. For ED, the MaxED bounds are generally set to $\lfloor 2^{0.1O} \rfloor$ and $\lfloor 2^{0.2O} \rfloor$. 
All methods are evaluated under the same constraint settings, and all reported approximate circuits satisfy the same target error bound.

\begin{table}[htbp]
\centering
\SetCaptionSpacing{0.03cm}{0.03cm}
\begin{threeparttable} 
\fontsize{6.8}{8.0}\selectfont
\setlength{\tabcolsep}{2.8pt}
\renewcommand{\arraystretch}{0.82}
\caption{Experimental benchmarks. Area and delay are measured by mapping the Compact AIGs into the ASAP 7nm library. (S: Size, D: Depth)}
\label{tab:experimental_benchmarks}
\begin{tabular}{@{} clc | rr | rr | rr | rr @{}} 
\toprule
\multirow{3}{*}{\begin{tabular}[c]{@{}c@{}}Benchmark\\ suite\end{tabular}} & \multirow{3}{*}{Circuit} & \multirow{3}{*}{PI/PO} & \multicolumn{6}{c|}{AIG} & \multicolumn{2}{c}{Gate netlist} \\
\cmidrule(lr){4-9} \cmidrule(l){10-11}
& & & \multicolumn{2}{c|}{Naive} & \multicolumn{2}{c|}{Compact} & \multicolumn{2}{c|}{Flattened} & Area & Delay \\
\cmidrule(lr){4-5} \cmidrule(lr){6-7} \cmidrule(lr){8-9}
& & & S & D & S & D & S & D & ($\mu m^2$) & (ps) \\
\midrule
\multirow{4}{*}{\begin{tabular}[c]{@{}c@{}}Used in\\ MECALS\end{tabular}} 
& mac      & 12/8    & 145  & 20  & 117  & 21  & 201   & 16  & 105.6  & 460.70  \\
& mult8    & 16/16   & 470  & 44  & 423  & 50  & 790   & 32  & 361.6  & 1068.71 \\
& mult16   & 32/32   & 2033 & 41  & 2000 & 64  & 2886  & 32  & 1706.4 & 1357.22 \\
& mult32   & 64/64   & 8340 & 53  & 8226 & 118 & 11586 & 41  & 7044.2 & 2506.34 \\
\midrule
\multirow{5}{*}{\begin{tabular}[c]{@{}c@{}}EPFL\\ arithmetic\tnote{*}\end{tabular}} 
& square8  & 8/16    & 235  & 28  & 188  & 26  & 285   & 15  & 153.3  & 586.72  \\
& log2\_8  & 8/8     & 3231 & 324 & 2346 & 289 & 5298  & 139 & 1635.0 & 6130.57 \\
& adder32  & 64/32   & 250  & 63  & 219  & 64  & 362   & 15  & 202.2  & 1452.66 \\
& adder64  & 128/64  & 506  & 127 & 443  & 128 & 879   & 26  & 411.2  & 2976.26 \\
& adder128 & 256/129 & 1020 & 255 & 892  & 256 & 1441  & 46  & 831.7  & 6062.91 \\
\midrule
\multirow{6}{*}{ISCAS85} 
& c880     & 60/26   & 330  & 30  & 300  & 27  & 385   & 14  & 227.2  & 556.38  \\
& c1908    & 33/25   & 421  & 33  & 360  & 26  & 540   & 20  & 283.2  & 558.72  \\
& c2670    & 233/140 & 718  & 22  & 543  & 23  & 713   & 14  & 458.9  & 470.71  \\
& c3540    & 50/22   & 1047 & 45  & 926  & 35  & 1159  & 26  & 600.6  & 855.73  \\
& c5315    & 178/123 & 1811 & 37  & 1280 & 32  & 1758  & 20  & 847.0  & 560.32  \\
& c7552    & 207/108 & 2192 & 30  & 1323 & 58  & 2093  & 19  & 1043.0 & 1427.98 \\
\bottomrule
\end{tabular}
\begin{tablenotes}
    \item[*] Benchmarks with reduced bit-widths (e.g., 8/32/64-bit) are derived from the original EPFL circuits by removing POs and performing structural strashing.
\end{tablenotes}
\end{threeparttable}
\vspace{-4mm}
\end{table}
\begin{table*}[t]
\centering

\caption{Comparison of E-ALS with baselines under the same error constraints. Subscripts $_{\mathrm{C}}$, $_{\mathrm{N}}$, and $_{\mathrm{F}}$ denote compact, naive, and flattened starts. \textbf{A}, \textbf{D}, and \textbf{R} denote Area (\%), Delay (\%), and Runtime (s). \textbf{MEC}, \textbf{SIM}, and \textbf{SIM+R} denote MECALS, SIM-ALS, and SIM-ALS+RESUB. \textbf{\textcolor{morandigreen}{Bold green}} marks the row-wise best Area achieved by E-ALS; $^{\dagger}$ and $^{\ddagger}$ mark the local best Areas within SIM and SIM+R. A dash indicates that no result was obtained within 48 hours. Means are computed over solved instances.}
\label{tab:ppa_compare_six_methods}

\vspace{-4mm}

\begingroup

\setlength{\tabcolsep}{2.5pt}

\renewcommand{\arraystretch}{0.86}

\setlength{\aboverulesep}{0.15ex}
\setlength{\belowrulesep}{0.15ex}
\setlength{\cmidrulesep}{0.10ex}

\resizebox{\textwidth}{!}{%
\begin{tabular}{l *{24}{c}}
\toprule
& \multicolumn{3}{c}{MEC$_{\mathrm{C}}$} & \multicolumn{3}{c}{SIM$_{\mathrm{C}}$} & \multicolumn{3}{c}{SIM$_{\mathrm{N}}$} & \multicolumn{3}{c}{SIM$_{\mathrm{F}}$} & \multicolumn{3}{c}{SIM+R$_{\mathrm{C}}$} & \multicolumn{3}{c}{SIM+R$_{\mathrm{N}}$} & \multicolumn{3}{c}{SIM+R$_{\mathrm{F}}$} & \multicolumn{3}{c}{E-ALS$_{\mathrm{N}}$} \\
\cmidrule(lr){2-4}\cmidrule(lr){5-7}\cmidrule(lr){8-10}\cmidrule(lr){11-13}\cmidrule(lr){14-16}\cmidrule(lr){17-19}\cmidrule(lr){20-22}\cmidrule(lr){23-25}
\multicolumn{1}{l|}{Circuit} & A & D & \multicolumn{1}{c|}{R} & A & D & \multicolumn{1}{c|}{R} & A & D & \multicolumn{1}{c|}{R} & A & D & \multicolumn{1}{c|}{R} & A & D & \multicolumn{1}{c|}{R} & A & D & \multicolumn{1}{c|}{R} & A & D & \multicolumn{1}{c|}{R} & A & D & R \\
\midrule
\multicolumn{1}{l|}{c880(HD2)} & 89.01 & 103.54 & \multicolumn{1}{c|}{1.42} & \textbf{84.08}$^{\dagger}$ & 87.63 & \multicolumn{1}{c|}{20.38} & 87.39 & 82.16 & \multicolumn{1}{c|}{20.41} & 85.85 & 92.26 & \multicolumn{1}{c|}{14.99} & 84.08 & 87.63 & \multicolumn{1}{c|}{18.96} & \textbf{83.87}$^{\ddagger}$ & 79.58 & \multicolumn{1}{c|}{27.11} & 85.85 & 92.26 & \multicolumn{1}{c|}{14.32} & \textcolor{morandigreen}{\textbf{76.83}} & 83.44 & 21.43 \\
\multicolumn{1}{l|}{c880(HD5)} & 85.70 & 97.28 & \multicolumn{1}{c|}{4.51} & 65.70 & 75.55 & \multicolumn{1}{c|}{37.36} & 63.87 & 89.58 & \multicolumn{1}{c|}{36.97} & \textbf{60.28}$^{\dagger}$ & 57.92 & \multicolumn{1}{c|}{38.64} & 65.70 & 75.55 & \multicolumn{1}{c|}{27.61} & 64.30 & 79.21 & \multicolumn{1}{c|}{46.01} & \textbf{60.28}$^{\ddagger}$ & 57.92 & \multicolumn{1}{c|}{36.71} & 63.45 & 75.88 & 26.96 \\
\multicolumn{1}{l|}{c1908(HD2)} & 92.32 & 102.17 & \multicolumn{1}{c|}{9.49} & \textbf{62.99}$^{\dagger}$ & 43.79 & \multicolumn{1}{c|}{18.24} & 67.80 & 79.38 & \multicolumn{1}{c|}{14.12} & 77.51 & 92.65 & \multicolumn{1}{c|}{35.25} & \textbf{63.84}$^{\ddagger}$ & 51.00 & \multicolumn{1}{c|}{15.50} & 67.57 & 81.25 & \multicolumn{1}{c|}{13.86} & 77.12 & 90.77 & \multicolumn{1}{c|}{33.81} & 65.48 & 86.95 & 30.40 \\
\multicolumn{1}{l|}{c1908(HD5)} & 60.85 & 78.76 & \multicolumn{1}{c|}{138.05} & 41.92 & 31.99 & \multicolumn{1}{c|}{42.35} & 44.29 & 32.62 & \multicolumn{1}{c|}{42.73} & \textbf{38.70}$^{\dagger}$ & 34.70 & \multicolumn{1}{c|}{55.51} & 41.92 & 31.99 & \multicolumn{1}{c|}{34.89} & 44.29 & 32.62 & \multicolumn{1}{c|}{47.21} & \textbf{38.70}$^{\ddagger}$ & 34.70 & \multicolumn{1}{c|}{56.46} & 43.11 & 33.48 & 36.72 \\
\multicolumn{1}{l|}{c2670(HD14)} & 77.89 & 93.55 & \multicolumn{1}{c|}{58.58} & 38.84 & 62.84 & \multicolumn{1}{c|}{285.44} & \textbf{37.66}$^{\dagger}$ & 60.80 & \multicolumn{1}{c|}{318.64} & 39.61 & 64.32 & \multicolumn{1}{c|}{425.55} & 38.84 & 62.84 & \multicolumn{1}{c|}{319.63} & \textbf{37.55}$^{\ddagger}$ & 58.93 & \multicolumn{1}{c|}{414.10} & 38.74 & 62.57 & \multicolumn{1}{c|}{408.94} & \textcolor{morandigreen}{\textbf{37.41}} & 39.92 & 83.41 \\
\multicolumn{1}{l|}{c2670(HD28)} & 36.75 & 59.24 & \multicolumn{1}{c|}{82.99} & 24.44 & 18.13 & \multicolumn{1}{c|}{228.06} & 26.60 & 23.30 & \multicolumn{1}{c|}{343.23} & \textbf{23.22}$^{\dagger}$ & 20.49 & \multicolumn{1}{c|}{402.84} & 24.44 & 18.13 & \multicolumn{1}{c|}{278.73} & 26.08 & 23.30 & \multicolumn{1}{c|}{394.39} & \textbf{23.22}$^{\ddagger}$ & 20.49 & \multicolumn{1}{c|}{394.35} & 29.99 & 24.43 & 62.80 \\
\multicolumn{1}{l|}{c3540(HD2)} & 92.43 & 92.35 & \multicolumn{1}{c|}{136.33} & 90.14 & 97.44 & \multicolumn{1}{c|}{13.02} & \textbf{89.10}$^{\dagger}$ & 94.97 & \multicolumn{1}{c|}{13.44} & 91.77 & 86.37 & \multicolumn{1}{c|}{22.46} & \textbf{89.90}$^{\ddagger}$ & 94.98 & \multicolumn{1}{c|}{16.89} & 90.14 & 92.41 & \multicolumn{1}{c|}{17.77} & 93.71 & 88.77 & \multicolumn{1}{c|}{24.51} & \textcolor{morandigreen}{\textbf{88.04}} & 101.26 & 50.83 \\
\multicolumn{1}{l|}{c3540(HD4)} & 90.30 & 91.53 & \multicolumn{1}{c|}{247.43} & 86.73 & 86.71 & \multicolumn{1}{c|}{54.53} & 85.22 & 80.82 & \multicolumn{1}{c|}{59.01} & \textbf{84.60}$^{\dagger}$ & 72.82 & \multicolumn{1}{c|}{71.84} & 86.95 & 86.27 & \multicolumn{1}{c|}{80.14} & \textbf{83.22}$^{\ddagger}$ & 80.83 & \multicolumn{1}{c|}{73.36} & 85.22 & 72.36 & \multicolumn{1}{c|}{93.72} & \textcolor{morandigreen}{\textbf{76.77}} & 67.96 & 148.81 \\
\multicolumn{1}{l|}{c5315(HD12)} & 94.69 & 96.59 & \multicolumn{1}{c|}{378.47} & 84.44 & 86.61 & \multicolumn{1}{c|}{193.62} & \textbf{72.89}$^{\dagger}$ & 88.08 & \multicolumn{1}{c|}{280.84} & \textbf{72.89}$^{\dagger}$ & 83.09 & \multicolumn{1}{c|}{294.14} & 83.64 & 88.75 & \multicolumn{1}{c|}{256.94} & \textbf{72.10}$^{\ddagger}$ & 81.48 & \multicolumn{1}{c|}{581.73} & 72.46 & 83.08 & \multicolumn{1}{c|}{364.77} & \textcolor{morandigreen}{\textbf{61.22}} & 97.94 & 276.92 \\
\multicolumn{1}{l|}{c5315(HD24)} & 85.44 & 97.17 & \multicolumn{1}{c|}{2341.79} & \textbf{54.61}$^{\dagger}$ & 79.02 & \multicolumn{1}{c|}{307.96} & 63.64 & 81.92 & \multicolumn{1}{c|}{281.26} & 60.96 & 77.34 & \multicolumn{1}{c|}{373.92} & \textbf{53.68}$^{\ddagger}$ & 75.28 & \multicolumn{1}{c|}{381.90} & 63.45 & 80.29 & \multicolumn{1}{c|}{696.81} & 60.29 & 76.40 & \multicolumn{1}{c|}{451.00} & 56.04 & 89.49 & 263.53 \\
\multicolumn{1}{l|}{c7552(HD10)} & 85.06 & 103.21 & \multicolumn{1}{c|}{601.25} & 37.21 & 95.52 & \multicolumn{1}{c|}{231.47} & 41.28 & 91.97 & \multicolumn{1}{c|}{269.09} & \textbf{34.64}$^{\dagger}$ & 81.78 & \multicolumn{1}{c|}{234.65} & 37.21 & 95.52 & \multicolumn{1}{c|}{285.60} & 40.24 & 96.60 & \multicolumn{1}{c|}{734.40} & \textbf{34.51}$^{\ddagger}$ & 75.46 & \multicolumn{1}{c|}{283.63} & \textcolor{morandigreen}{\textbf{32.95}} & 108.62 & 255.94 \\
\multicolumn{1}{l|}{c7552(HD21)} & 73.34 & 100.06 & \multicolumn{1}{c|}{2478.68} & \textbf{27.83}$^{\dagger}$ & 81.95 & \multicolumn{1}{c|}{303.74} & 30.17 & 88.02 & \multicolumn{1}{c|}{288.28} & 29.59 & 52.66 & \multicolumn{1}{c|}{405.02} & 28.04 & 82.14 & \multicolumn{1}{c|}{353.58} & 30.13 & 81.09 & \multicolumn{1}{c|}{594.02} & \textbf{27.69}$^{\ddagger}$ & 51.58 & \multicolumn{1}{c|}{450.18} & 27.98 & 88.29 & 312.08 \\
\midrule
\multicolumn{1}{l|}{\textbf{HD Mean}} & 80.32 & 92.96 & \multicolumn{1}{c|}{-} & 58.25 & 70.60 & \multicolumn{1}{c|}{-} & 59.16 & 74.47 & \multicolumn{1}{c|}{-} & 58.30 & 68.03 & \multicolumn{1}{c|}{-} & 58.19 & 70.84 & \multicolumn{1}{c|}{-} & 58.58 & 72.30 & \multicolumn{1}{c|}{-} & 58.15 & 67.20 & \multicolumn{1}{c|}{-} & 54.94 & 74.80 & - \\
\midrule
\multicolumn{1}{l|}{mac(ED1)} & 89.09 & 87.36 & \multicolumn{1}{c|}{2.33} & 94.39 & 97.77 & \multicolumn{1}{c|}{0.20} & \textbf{89.24}$^{\dagger}$ & 98.03 & \multicolumn{1}{c|}{0.45} & 89.70 & 84.62 & \multicolumn{1}{c|}{0.84} & 94.39 & 97.77 & \multicolumn{1}{c|}{0.27} & 89.24 & 98.03 & \multicolumn{1}{c|}{0.54} & \textbf{88.79}$^{\ddagger}$ & 89.95 & \multicolumn{1}{c|}{1.32} & \textcolor{morandigreen}{\textbf{82.88}} & 91.47 & 11.56 \\
\multicolumn{1}{l|}{mac(ED2)} & 86.97 & 99.09 & \multicolumn{1}{c|}{1.65} & \textbf{86.67}$^{\dagger}$ & 85.79 & \multicolumn{1}{c|}{0.34} & 87.42 & 98.00 & \multicolumn{1}{c|}{0.71} & 88.64 & 88.29 & \multicolumn{1}{c|}{1.48} & \textbf{86.67}$^{\ddagger}$ & 85.79 & \multicolumn{1}{c|}{0.45} & 87.42 & 98.00 & \multicolumn{1}{c|}{0.99} & 87.73 & 84.52 & \multicolumn{1}{c|}{2.04} & \textcolor{morandigreen}{\textbf{77.58}} & 92.81 & 18.35 \\
\multicolumn{1}{l|}{square8(ED1)} & 98.02 & 96.97 & \multicolumn{1}{c|}{1.54} & 96.14 & 92.32 & \multicolumn{1}{c|}{0.49} & \textbf{93.01}$^{\dagger}$ & 87.98 & \multicolumn{1}{c|}{0.81} & 96.66 & 89.68 & \multicolumn{1}{c|}{1.38} & 96.14 & 92.32 & \multicolumn{1}{c|}{0.61} & \textbf{93.42}$^{\ddagger}$ & 88.22 & \multicolumn{1}{c|}{0.97} & 98.43 & 78.86 & \multicolumn{1}{c|}{3.27} & \textcolor{morandigreen}{\textbf{90.08}} & 86.45 & 26.15 \\
\multicolumn{1}{l|}{square8(ED3)} & 98.02 & 96.97 & \multicolumn{1}{c|}{2.33} & 96.14 & 92.32 & \multicolumn{1}{c|}{0.56} & \textbf{93.01}$^{\dagger}$ & 84.68 & \multicolumn{1}{c|}{0.62} & 94.47 & 85.57 & \multicolumn{1}{c|}{2.16} & 96.14 & 92.32 & \multicolumn{1}{c|}{0.63} & \textbf{93.42}$^{\ddagger}$ & 84.92 & \multicolumn{1}{c|}{0.84} & 100.63 & 74.87 & \multicolumn{1}{c|}{4.48} & \textcolor{morandigreen}{\textbf{89.14}} & 85.07 & 78.06 \\
\multicolumn{1}{l|}{log2\_8(ED1)} & 18.89 & 20.56 & \multicolumn{1}{c|}{36538.51} & \textbf{6.73}$^{\dagger}$ & 10.60 & \multicolumn{1}{c|}{42.36} & 7.64 & 10.15 & \multicolumn{1}{c|}{92.96} & 7.89 & 10.35 & \multicolumn{1}{c|}{100.89} & \textbf{6.73}$^{\ddagger}$ & 10.60 & \multicolumn{1}{c|}{43.06} & 7.64 & 10.15 & \multicolumn{1}{c|}{104.38} & 7.70 & 9.67 & \multicolumn{1}{c|}{121.76} & \textcolor{morandigreen}{\textbf{6.04}} & 8.92 & 102.40 \\
\multicolumn{1}{l|}{log2\_8(ED3)} & 7.66 & 9.43 & \multicolumn{1}{c|}{23234.12} & 6.28 & 9.86 & \multicolumn{1}{c|}{37.53} & \textbf{5.88}$^{\dagger}$ & 9.08 & \multicolumn{1}{c|}{108.67} & 6.26 & 10.24 & \multicolumn{1}{c|}{148.83} & 6.28 & 9.86 & \multicolumn{1}{c|}{53.39} & \textbf{6.08}$^{\ddagger}$ & 10.33 & \multicolumn{1}{c|}{122.75} & 6.10 & 9.02 & \multicolumn{1}{c|}{154.54} & \textcolor{morandigreen}{\textbf{5.60}} & 8.13 & 129.15 \\
\multicolumn{1}{l|}{adder32(ED9)} & 95.97 & 96.31 & \multicolumn{1}{c|}{93.14} & 95.17 & 94.36 & \multicolumn{1}{c|}{13.78} & 95.17 & 94.36 & \multicolumn{1}{c|}{10.38} & \textbf{92.88}$^{\dagger}$ & 92.00 & \multicolumn{1}{c|}{11.54} & 95.17 & 94.36 & \multicolumn{1}{c|}{13.74} & 95.17 & 94.36 & \multicolumn{1}{c|}{12.66} & \textbf{92.88}$^{\ddagger}$ & 92.00 & \multicolumn{1}{c|}{13.36} & \textcolor{morandigreen}{\textbf{80.06}} & 84.90 & 22.66 \\
\multicolumn{1}{l|}{adder32(ED84)} & 89.24 & 86.81 & \multicolumn{1}{c|}{178.38} & 86.47 & 88.51 & \multicolumn{1}{c|}{26.42} & 86.95 & 85.63 & \multicolumn{1}{c|}{27.59} & \textbf{84.65}$^{\dagger}$ & 81.86 & \multicolumn{1}{c|}{45.28} & 86.47 & 88.51 & \multicolumn{1}{c|}{21.59} & 86.95 & 85.63 & \multicolumn{1}{c|}{24.50} & \textbf{84.65}$^{\ddagger}$ & 81.86 & \multicolumn{1}{c|}{44.54} & \textcolor{morandigreen}{\textbf{76.11}} & 78.72 & 38.55 \\
\multicolumn{1}{l|}{adder64(ED84)} & 94.75 & 93.04 & \multicolumn{1}{c|}{3385.57} & \textbf{94.12}$^{\dagger}$ & 94.02 & \multicolumn{1}{c|}{57.78} & 94.40 & 94.28 & \multicolumn{1}{c|}{64.75} & 99.38 & 68.28 & \multicolumn{1}{c|}{58.00} & \textbf{93.89}$^{\ddagger}$ & 92.65 & \multicolumn{1}{c|}{74.00} & \textbf{93.89}$^{\ddagger}$ & 92.65 & \multicolumn{1}{c|}{79.21} & 96.61 & 69.71 & \multicolumn{1}{c|}{59.44} & \textcolor{morandigreen}{\textbf{79.18}} & 83.58 & 74.89 \\
\multicolumn{1}{l|}{adder64(ED7131)} & 92.14 & 89.82 & \multicolumn{1}{c|}{4858.65} & \textbf{87.32}$^{\dagger}$ & 88.44 & \multicolumn{1}{c|}{133.89} & 88.09 & 90.25 & \multicolumn{1}{c|}{152.43} & 92.26 & 64.57 & \multicolumn{1}{c|}{150.25} & \textbf{88.09}$^{\ddagger}$ & 90.25 & \multicolumn{1}{c|}{159.42} & \textbf{88.09}$^{\ddagger}$ & 90.25 & \multicolumn{1}{c|}{174.94} & 95.29 & 55.33 & \multicolumn{1}{c|}{167.87} & \textcolor{morandigreen}{\textbf{74.94}} & 75.37 & 170.28 \\
\multicolumn{1}{l|}{adder128(ED7600)} & 90.52 & 89.59 & \multicolumn{1}{c|}{91191.26} & 90.40 & 89.59 & \multicolumn{1}{c|}{114.60} & 90.40 & 89.59 & \multicolumn{1}{c|}{191.52} & \textbf{85.98}$^{\dagger}$ & 72.67 & \multicolumn{1}{c|}{179.03} & 90.40 & 89.59 & \multicolumn{1}{c|}{165.93} & 90.40 & 89.59 & \multicolumn{1}{c|}{197.81} & \textbf{87.26}$^{\ddagger}$ & 71.72 & \multicolumn{1}{c|}{277.00} & \textcolor{morandigreen}{\textbf{74.47}} & 79.44 & 278.03 \\
\multicolumn{1}{l|}{adder128(ED58e6)} & 80.30 & 79.48 & \multicolumn{1}{c|}{130176.85} & 80.17 & 79.49 & \multicolumn{1}{c|}{356.41} & 80.49 & 79.51 & \multicolumn{1}{c|}{453.23} & \textbf{76.11}$^{\dagger}$ & 63.74 & \multicolumn{1}{c|}{397.62} & 80.17 & 79.49 & \multicolumn{1}{c|}{393.73} & 80.49 & 79.51 & \multicolumn{1}{c|}{559.63} & \textbf{76.93}$^{\ddagger}$ & 63.41 & \multicolumn{1}{c|}{408.36} & \textcolor{morandigreen}{\textbf{66.08}} & 70.37 & 414.04 \\
\multicolumn{1}{l|}{mult8(ED3)} & 96.95 & 93.25 & \multicolumn{1}{c|}{104.60} & 95.80 & 93.25 & \multicolumn{1}{c|}{2.46} & \textbf{95.66}$^{\dagger}$ & 93.94 & \multicolumn{1}{c|}{1.99} & 98.01 & 87.66 & \multicolumn{1}{c|}{2.05} & 95.80 & 93.25 & \multicolumn{1}{c|}{2.15} & \textbf{93.72}$^{\ddagger}$ & 96.26 & \multicolumn{1}{c|}{3.29} & 97.65 & 87.43 & \multicolumn{1}{c|}{7.02} & \textcolor{morandigreen}{\textbf{92.83}} & 90.50 & 87.42 \\
\multicolumn{1}{l|}{mult8(ED9)} & 93.85 & 93.49 & \multicolumn{1}{c|}{230.31} & \textbf{93.01}$^{\dagger}$ & 93.49 & \multicolumn{1}{c|}{4.83} & 93.63 & 94.22 & \multicolumn{1}{c|}{6.49} & 95.53 & 90.52 & \multicolumn{1}{c|}{9.17} & 93.01 & 93.49 & \multicolumn{1}{c|}{5.32} & \textbf{91.68}$^{\ddagger}$ & 96.38 & \multicolumn{1}{c|}{10.12} & 95.40 & 89.57 & \multicolumn{1}{c|}{25.19} & \textcolor{morandigreen}{\textbf{89.34}} & 96.23 & 189.23 \\
\multicolumn{1}{l|}{mult16(ED9)} & 95.69 & 95.00 & \multicolumn{1}{c|}{18079.44} & 95.67 & 94.90 & \multicolumn{1}{c|}{15.33} & 96.53 & 98.37 & \multicolumn{1}{c|}{13.91} & \textbf{94.81}$^{\dagger}$ & 98.98 & \multicolumn{1}{c|}{27.88} & 95.71 & 94.82 & \multicolumn{1}{c|}{18.27} & 95.99 & 98.97 & \multicolumn{1}{c|}{21.82} & \textbf{93.66}$^{\ddagger}$ & 90.89 & \multicolumn{1}{c|}{45.66} & \textcolor{morandigreen}{\textbf{91.09}} & 122.26 & 426.93 \\
\multicolumn{1}{l|}{mult16(ED84)} & 91.22 & 79.52 & \multicolumn{1}{c|}{36086.81} & 90.97 & 81.68 & \multicolumn{1}{c|}{114.28} & 91.66 & 90.03 & \multicolumn{1}{c|}{198.17} & \textbf{89.66}$^{\dagger}$ & 85.96 & \multicolumn{1}{c|}{227.40} & 90.32 & 87.01 & \multicolumn{1}{c|}{144.76} & 91.69 & 90.04 & \multicolumn{1}{c|}{183.39} & \textbf{90.03}$^{\ddagger}$ & 83.64 & \multicolumn{1}{c|}{356.38} & \textcolor{morandigreen}{\textbf{87.79}} & 124.53 & 1100.57 \\
\multicolumn{1}{l|}{mult32(ED84)} & - & - & \multicolumn{1}{c|}{-} & 94.84 & 96.50 & \multicolumn{1}{c|}{731.23} & 94.99 & 96.97 & \multicolumn{1}{c|}{803.37} & \textbf{94.56}$^{\dagger}$ & 80.77 & \multicolumn{1}{c|}{1064.74} & 94.78 & 93.13 & \multicolumn{1}{c|}{748.97} & 94.99 & 96.97 & \multicolumn{1}{c|}{880.07} & \textbf{94.61}$^{\ddagger}$ & 78.06 & \multicolumn{1}{c|}{1297.28} & \textcolor{morandigreen}{\textbf{92.42}} & 118.93 & 4013.64 \\
\multicolumn{1}{l|}{mult32(ED7131)} & - & - & \multicolumn{1}{c|}{-} & \textbf{90.76}$^{\dagger}$ & 90.46 & \multicolumn{1}{c|}{574.27} & 91.06 & 91.73 & \multicolumn{1}{c|}{2231.89} & 90.77 & 72.51 & \multicolumn{1}{c|}{5195.53} & 90.71 & 87.09 & \multicolumn{1}{c|}{694.04} & 90.80 & 91.30 & \multicolumn{1}{c|}{2040.26} & \textbf{90.67}$^{\ddagger}$ & 74.84 & \multicolumn{1}{c|}{5590.25} & \textcolor{morandigreen}{\textbf{88.81}} & 120.66 & 6474.26 \\
\midrule
\multicolumn{1}{l|}{\textbf{ED Mean}} & 82.45 & 81.67 & \multicolumn{1}{c|}{-} & 82.28 & 81.85 & \multicolumn{1}{c|}{-} & 81.96 & 82.60 & \multicolumn{1}{c|}{-} & 82.12 & 73.79 & \multicolumn{1}{c|}{-} & 82.27 & 81.80 & \multicolumn{1}{c|}{-} & 81.73 & 82.86 & \multicolumn{1}{c|}{-} & 82.50 & 71.41 & \multicolumn{1}{c|}{-} & 74.69 & 84.35 & - \\
\bottomrule
\end{tabular}%
}

\endgroup

\vspace{-4mm}
\end{table*}

The main evaluation metrics are the final area and delay, normalized to the exact gate-level netlist of each benchmark. To obtain them, all resulting approximate netlists are post-processed by a unified flow: \texttt{"2$\times$compress2rs"}, followed by technology mapping using \texttt{ABC "dch+amap"}. 

Our framework is implemented in Rust and C, integrating the e-graph rewriting engine with ABC-based ALS routines. Experiments are conducted on a server with two Intel Xeon Platinum 8358P CPUs. For all methods, the sampled pattern count is fixed to 1024, and the ALS procedure examines at most $k=100$ LACs in each round. In E-ALS, rewriting uses four progressive expansion stages with node-expansion caps of $2.5\times$, $5\times$, $7.5\times$, and $10\times$ the initial graph size. Simulated-annealing extraction uses 10 iterations, a batch size of 16, and a perturbation ratio of 10\%. To enable budget-guided structural refinement, E-ALS uses a multi-stage budget schedule during optimization, with up to $T=6$ stages in our experiments. Since the compared ALS flows are stochastic, each stochastic method is repeated 3 times, and the best final-area result is reported together with its corresponding delay and runtime.

\subsection{Comparison with Structural ALS Baselines}

We compare E-ALS with representative structural ALS baselines, including MECALS~\cite{meng2023mecals}, SIM-ALS~\cite{meng2026simulation}, and SIM-ALS+RESUB~\cite{meng2024efficient}. Here, SIM-ALS serves as the strong baseline, while SIM-ALS+RESUB tests whether simply strengthening the local approximation model is sufficient. For SIM-ALS and SIM-ALS+RESUB, we report results from compact, naive, and flattened starts, denoted by $_{\mathrm{C}}$, $_{\mathrm{N}}$, and $_{\mathrm{F}}$, respectively. MECALS follows its standard compact-start setup, whereas E-ALS is evaluated from the naive start, since it explicitly avoids committing to an exact-oriented structural prior before approximation-aware exploration. 

Table~\ref{tab:ppa_compare_six_methods} reports the normalized final area, delay, and runtime under the same error constraints. Overall, E-ALS achieves the lowest average normalized final area in both error models, reaching 54.94\% in HD and 74.69\% in ED. This advantage is also evident in representative cases: on \texttt{c5315(HD12)}, E-ALS$_{\mathrm{N}}$ achieves a normalized final area of 61.22\%, while all SIM and SIM-ALS+RESUB variants remain above 72\%; on \texttt{adder32(ED9)}, it achieves 80.06\%, whereas all SIM and SIM-ALS+RESUB variants remain above 92\%. Beyond these aggregate improvements, the results reveal two clear trends. First, the initial exact structure is not a neutral starting point, since the relative advantage among compact, naive, and flattened starts frequently reverses across benchmarks and error settings. Second, simply strengthening the local approximation model yields only limited additional gains under practical runtime budgets, whereas further increasing its expressiveness tends to incur substantial scalability costs. Overall, the results show that approximation-aware structural exploration can uncover substantially better ALS starting structures than fixed exact starts.

Overall, these results support our central claim that the key limitation of structural ALS lies not only in the local approximation operator, but also in the structural substrate itself. Accordingly, E-ALS delivers substantially stronger area reduction than either relying on a fixed exact start or only strengthening the local approximation model. In overall runtime, E-ALS remains broadly comparable to prior structural ALS methods. When its total runtime becomes higher in some cases, the increase stems from two factors. First, E-ALS incurs the cumulative cost of exploring multiple equivalent structures across progressive budget stages (up to 6). Second, the rewritten netlists often exhibit SAT-verification behavior that differs markedly from that of the original structures; empirically, they tend to slow down exact checking, especially near the error-budget boundary, where verification may repeatedly hit the SAT runtime limit. Reducing the amount of structural exploration while preserving most of the area benefit, and better understanding how rewritten structures affect exact verification cost, are important directions for future work. A detailed runtime breakdown is given in the runtime analysis subsection.

\subsection{Ablation Study of Core Design Components}

To isolate the contribution of the two core design components of E-ALS, we perform a budget-controlled two-factor ablation on Function-Reduced Saturation (FRS) and the ALS-coupled surrogate objective. This yields four variants: \textbf{Full}, \textbf{w/o FRS}, \textbf{w/o ApproxObj}, and \textbf{w/o Both}. Here, \textbf{w/o ApproxObj} keeps the same rewritten e-graph, simulated-annealing search, and budget-guided refinement flow, but replaces the surrogate objective with the exact AIG size of the extracted structure, in order to test whether exact compactness is already sufficient for structure selection within the same framework. In contrast, \textbf{w/o FRS} keeps the same surrogate objective but removes FRS, in order to test whether a good ranking signal alone is sufficient when the equivalence space remains more fragmented. All other settings are kept unchanged.

\begin{table}[t]
\centering
\SetCaptionSpacing{0.03cm}{0.03cm}
\fontsize{6.8}{8.0}\selectfont
\setlength{\tabcolsep}{2.8pt}
\renewcommand{\arraystretch}{0.8}
\caption{Ablation on the two core components of E-ALS. Lower Area is better. $\Delta$ denotes the area degradation relative to Full in percentage points.}
\label{tab:rq2_ablation}
\setlength{\tabcolsep}{4pt} 
\resizebox{1\columnwidth}{!}{%
\begin{tabular}{l ccc ccc} 
\toprule
& \multicolumn{3}{c}{\textbf{HD Benchmarks}} & \multicolumn{3}{c}{\textbf{ED Benchmarks}} \\
\cmidrule(lr){2-4} \cmidrule(lr){5-7}
Variant & Area ($\downarrow$) & $\Delta$ & Delay & Area ($\downarrow$) & $\Delta$ & Delay \\
\midrule
\textbf{Full (E-ALS)} & \textbf{54.94} & -- & 74.80 & \textbf{74.69} & -- & 84.35 \\
\midrule
\quad w/o FRS         & 59.25 & +4.31 & 79.59 & 77.06 & +2.37 & 85.80 \\
\quad w/o ApproxObj   & 60.47 & +5.53 & 84.25 & 84.66 & +9.97 & 94.56 \\
\quad w/o Both        & 60.89 & +5.95 & 86.91 & 85.03 & +10.34 & 97.14 \\
\bottomrule
\end{tabular}%
}
\vspace{-5mm}
\end{table}

Table~\ref{tab:rq2_ablation} shows that \textbf{Full} achieves the best final area under both HD and ED constraints. Among the two components, the surrogate objective has the stronger impact: replacing it with exact AIG size increases the mean final area from 54.94\% to 60.47\% under HD and from 74.69\% to 84.66\% under ED, even though the same structural search framework is retained. This indicates that the gain of E-ALS does not arise merely from structural exploration itself, but from ranking candidate structures by downstream approximation potential rather than exact compactness, and also confirms that a more compact exact starting structure does not necessarily lead to a better approximate result. FRS contributes independently as well: disabling it while keeping the same surrogate objective still degrades the mean final area, and removing both components yields the worst result. Overall, these results show that E-ALS benefits from both approximation-aware ranking and a less fragmented, more extraction-friendly equivalence space.



\subsection{Budget-Guided Structural Refinement and Stage-Wise Oracle Analysis}

\begin{figure*}[tb!]
  \centering
  \SetCaptionSpacing{0.10cm}{0.10cm}
  \includegraphics[width=1.0\linewidth]{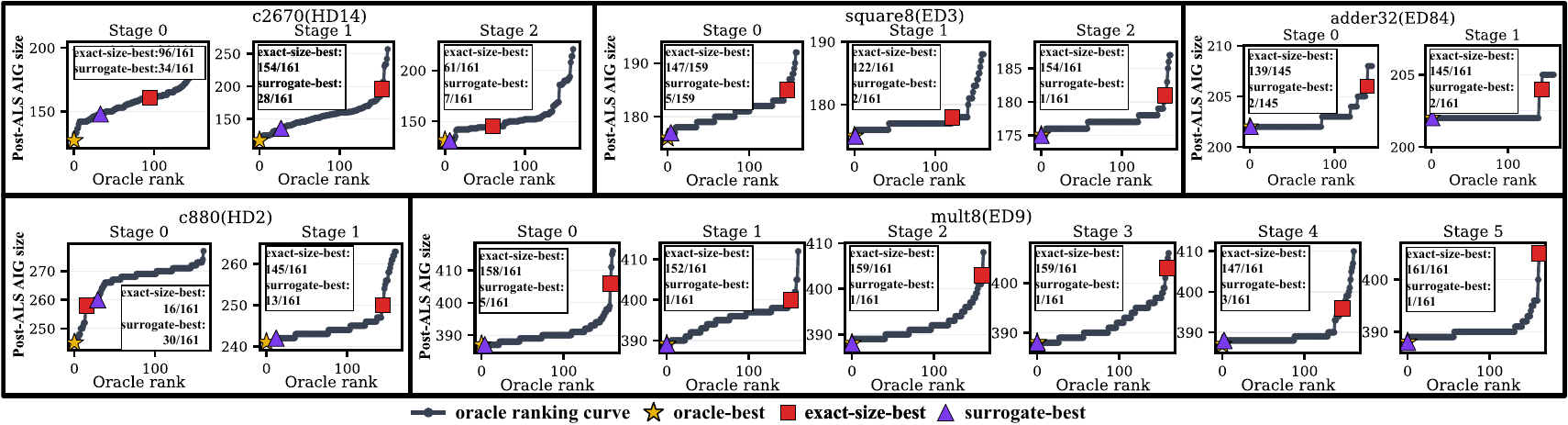} 
  \caption{Oracle-based stage-wise ranking analysis on five representative benchmark-constraint groups (two HD and three ED cases). In each subplot, candidate structures encountered at one budget stage are ranked by their oracle post-ALS AIG size. The surrogate-best candidate stays consistently much closer to the oracle optimum than the exact-size-best candidate.}
  \label{fig:E3}
  \vspace{-4mm}
\end{figure*}

\begin{figure}[tb!]
  \centering
  \SetCaptionSpacing{0.10cm}{0.10cm}
  \includegraphics[width=1.0\linewidth]{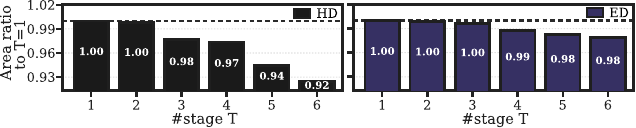} 
  \caption{Average final-area ratio versus the number of budget stages $T$, normalized to $T=1$. Increasing $T$ benefits HD more than ED, whose gains saturate quickly.}
  \label{fig:E3_2}
  \vspace{-4mm}
\end{figure}

We evaluate budget-guided structural refinement both at the end-to-end level and through stage-wise oracle analysis. As shown in Fig.~\ref{fig:E3_2}, increasing the number of budget stages improves final quality in both error models, with a more pronounced effect under HD. In particular, later budget stages continue to bring clear gains under HD, while still providing measurable improvement under ED. This indicates that budget-guided refinement is beneficial in both settings, although its impact is stronger under HD.

The stage-wise oracle analysis in Fig.~\ref{fig:E3} further explains this benefit. At each budget stage, all candidate structures encountered during search are ranked by their oracle post-ALS AIG size, allowing us to compare the oracle-best, exact-size-best, and surrogate-best candidates on the same candidate pool. The exact-size-best candidate is often far from the oracle optimum, whereas the surrogate-best candidate remains much closer throughout the refinement process. In particular, on \texttt{c2670(HD14)} and \texttt{c880(HD2)}, the surrogate-best candidate becomes noticeably closer to the oracle-best structure as the budget stage progresses. This indicates that exact structural compactness is poorly aligned with downstream ALS quality on the actual structures encountered during search, while the surrogate-based ranking has the potential to become more informative as optimization proceeds. Therefore, budget-guided refinement is effective not only because it revisits structure selection, but also because later refinement stages enable more accurate structure ranking under richer error information.

\subsection{Runtime Breakdown and Parallel Screening Efficiency}

\begin{figure}[tb!]
  \centering
  \SetCaptionSpacing{0.10cm}{0.10cm}
  \includegraphics[width=0.82\linewidth]{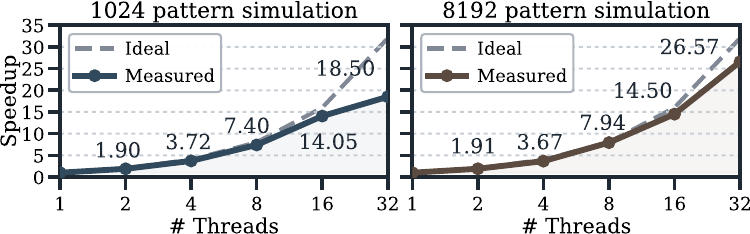} 
  \caption{Thread scalability of parallel raw-LAC screening on \texttt{mult32(ED7131)} with a cap of 100000 raw LACs.}
  \label{fig:E4_1}
  \vspace{-4mm}
\end{figure}

We analyze the runtime of E-ALS from two perspectives: the scalability of parallel raw-LAC screening and the end-to-end runtime breakdown of the overall flow. As shown in Fig.~\ref{fig:E4_1}, raw-LAC screening scales well with thread count, reaching 14.05$\times$ speedup at 16 threads and 18.50$\times$ at 32 threads. This confirms that the target-centric task decomposition and dynamic load balancing effectively mitigate one of the main computational bottlenecks of E-ALS, although the speedup becomes sublinear at higher thread counts.

\begin{figure}[tb!]
  \centering
  \SetCaptionSpacing{0.10cm}{0.10cm}
  \includegraphics[width=0.85\linewidth]{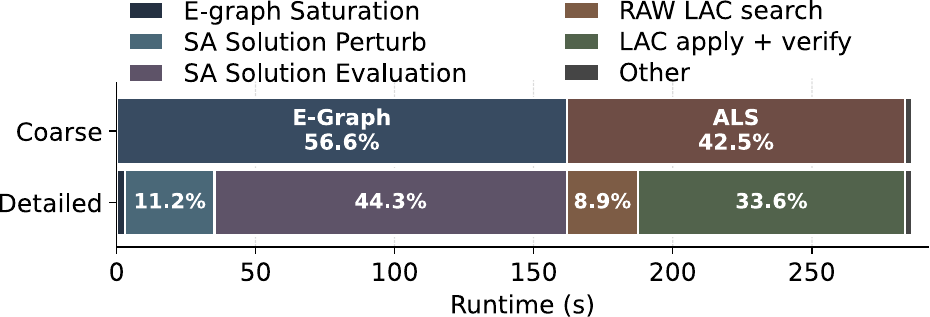} 
  \caption{Runtime breakdown of an E-ALS run on \texttt{adder128(ED7600)} under parallel evaluation. Each raw-LAC screening uses 2 workers, while simulated annealing evaluates 16 candidates in parallel, yielding up to 32 concurrent evaluation workers.}
  \label{fig:E4_2}
  \vspace{-4mm}
\end{figure}

Fig.~\ref{fig:E4_2} shows that the runtime of E-ALS is dominated by repeated structure evaluation and ALS execution, rather than by rewriting alone. In the representative \texttt{adder128(ED7600)} run, the E-graph side accounts for 56.6\% of the total runtime, while the ALS side accounts for 42.5\%. At finer granularity, SA solution evaluation is the largest component (44.3\%), followed by LAC apply-and-verify (33.6\%) and explicit raw-LAC screening (8.9\%). Because SA solution evaluation repeatedly invokes the same screening machinery, the benefit of parallel raw-LAC screening extends beyond its explicit 8.9\% runtime share. Together with the area gains, these results indicate that the additional runtime of E-ALS is primarily spent on approximation-aware structural evaluation, which is central to its stronger area reduction, and that this cost is substantially alleviated by effective parallel screening.

\section{Conclusion}\label{conclusion}
This paper presents \textbf{E-ALS}, an e-graph-based framework for approximate logic synthesis that addresses \emph{structural bias} in structural ALS. By combining equivalence-enhanced structural search, an ALS-coupled surrogate for approximation-aware ranking, and budget-guided structural refinement, E-ALS identifies better exact structures for downstream approximation under maximum-error constraints. Experiments on arithmetic and logic benchmarks show that E-ALS outperforms state-of-the-art baselines, demonstrating that approximation-aware exploration over equivalent structures can uncover hidden opportunities beyond conventional exact-oriented optimization.
\newpage

\balance
\bibliographystyle{ACM-Reference-Format}
\bibliography{bibfile}

\end{document}